\documentclass{iau} 
\usepackage{graphicx}

\title[Snowlines in protostellar environments] 
{Imaging the water snowline in protostellar envelopes}

\author[Merel L.R. van 't Hoff]   
{Merel L.R. van 't Hoff$^1$}

\affiliation{$^1$Leiden Observatory, Leiden University, PO box 9513,
NL-2300 RA, Leiden, the Netherlands \\ email: {\tt vthoff@strw.leidenuniv.nl}}

\pubyear{2017}
\volume{332}  
\jname{Astrochemistry VII -- Through the Cosmos from Galaxies to Planets}
\editors{M. Cunningham, T. Millar \& Y. Aikawa, eds.}

\begin{document}

\maketitle


\begin{abstract} 
Determining the locations of the major snowlines in protostellar environments is crucial to fully understand the planet formation process and its outcome. Despite being located far enough from the central star to be spatially resolved with ALMA, the CO snowline remains difficult to detect directly in protoplanetary disks. Instead, its location can be derived from N$_2$H$^+$ emission, when chemical effects like photodissociation of CO and N$_2$ are taken into account. The water snowline is even harder to observe than that for CO, because in disks it is located only a few AU from the protostar, and from the ground only the less abundant isotopologue H$_2^{18}$O can be observed. Therefore, using an indirect chemical tracer, as done for CO, may be the best way to locate the water snowline. A good candidate tracer is HCO$^+$, which is expected to be particularly abundant when its main destructor, H$_2$O, is frozen out. Comparison of H$_2^{18}$O and H$^{13}$CO$^+$ emission toward the envelope of the Class 0 protostar IRAS2A shows that the emission from both molecules is spatially anticorrelated, providing a proof of concept that H$^{13}$CO$^+$ can indeed be used to trace the water snowline in systems where it cannot be imaged directly. 


\end{abstract}

       
\firstsection 
                     
\section{Introduction}

The formation of low-mass stars begins with the collapse of a dense core in a molecular cloud. To conserve angular momentum, the infalling material forms a disk around the protostar. Due to the ongoing accretion of material from the surrounding envelope through the disk onto the star, together with the launching of an outflow, the envelope dissipates over time. What remains is a pre-main sequence star surrounded by a protoplanetary disk that contains the gas and dust from which a planetary system may be forming. The composition of the planets is thus determined by the chemical structure of the protostellar environment. 

\begin{figure}[t]
\begin{center}
 \includegraphics[width=3.4in]{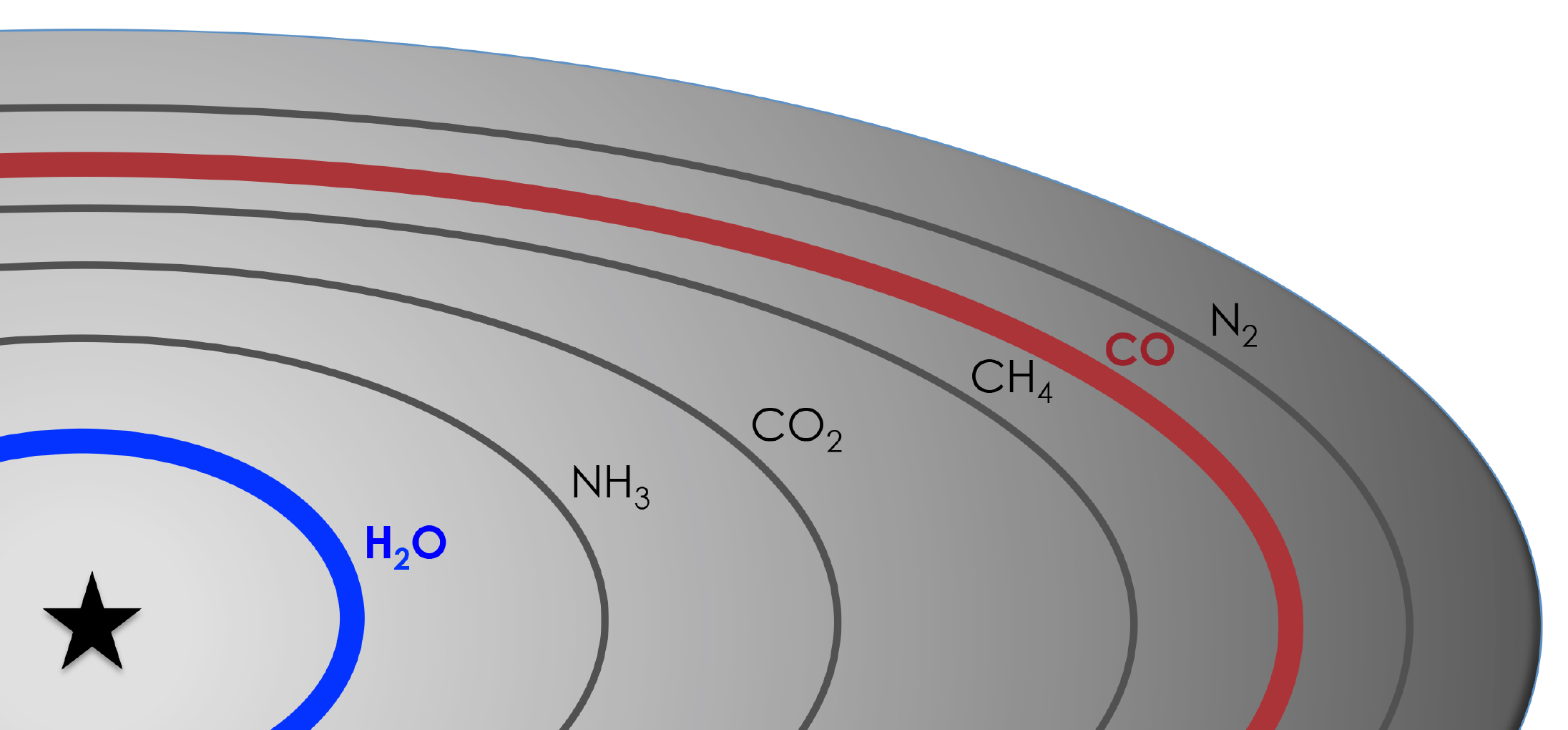} 
 \caption{Schematic overview of the major snowlines. Their relative positions are set by the species-dependent freeze-out temperature, while the absolute positions depend on the physical structure of each individual object. The CO and H$_2$O snowlines discussed here are highlighted.}
 \label{fig:Snowlines}
\end{center}
\end{figure}

A key aspect of protostellar chemistry are snowlines. A snowline marks the midplane radius at which a molecular species freezes out from the gas phase onto dust grains. The location of a snowline depends both on the species-dependent sublimation temperature and the physical structure of the protostellar envelope or protoplanetary disk (e.g., temperature and density; see Fig.~\ref{fig:Snowlines}). The selective freeze out of major carbon or oxygen carrying species at different snowlines causes the elemental C/O-ratio of the planet forming material to vary with radial distance from the star (\cite{Oberg2011}; \cite{Oberg2016}). The bulk composition of planets may therefore be regulated by their formation location with respect to the major snowlines (e.g., \cite{Madhusudhan2014}; \cite{Walsh2015}; \cite{Eistrup2016}). Moreover, the growth of dust particles, and thus the planet formation efficiency is thought to be enhanced in these freeze-out zones, for example due to the increase in solid density (e.g., \cite{Stevenson1988}; \cite{Ros2013}; \cite{Schoonenberg2017}). Snowlines thus play a crucial role in the formation and composition of planets. Determining their locations is therefore key to fully understand the planet formation process and its outcome. 

\section{The CO snowline} 

Of the major snowlines, the CO snowline is particularly interesting because CO ice is the starting point for the formation of many complex organic molecules (e.g., \cite{Herbst2009}). In addition, due to the low sublimation temperature of CO ($\sim$20~K), the CO snowline is located relatively far away from the central star (10s--100 AU in protoplanetary disks; see Fig.~\ref{fig:Snowlines}). Although we are now able to spatially resolve this with ALMA, locating the CO snowline directly remains difficult. CO line emission is generally optically thick and does not reveal the cold disk midplane. The logical step in such situation would be to use optically thin isotopologues. However, even C$^{18}$O does not always offer a solution as is seen for TW Hya (\cite{Schwarz2016}): the CO abundance remains 1--2 orders of magnitude below the $10^{-4}$ ISM abundance far within the radius where the CO snowline is expected based on the disk temperature. Furthermore, a recent ALMA survey of all disks in the Lupus star forming region suggest that CO abundances below the ISM abundance may be common (\cite{Ansdell2016}; \cite{Miotello2017}), indicating that other processes than freeze-out may contribute to lowering the gas-phase CO abundance. In contrast to C$^{18}$O, the double isotopologue $^{13}$C$^{18}$O has recently been shown to reveal the snowline location in TW~Hya (\cite{Zhang2017}). However, the low $^{13}$C$^{18}$O abundance restricts observations to the very few nearest and brightest disks. 

\begin{figure}[t]
\begin{center}
 \includegraphics[width=\linewidth]{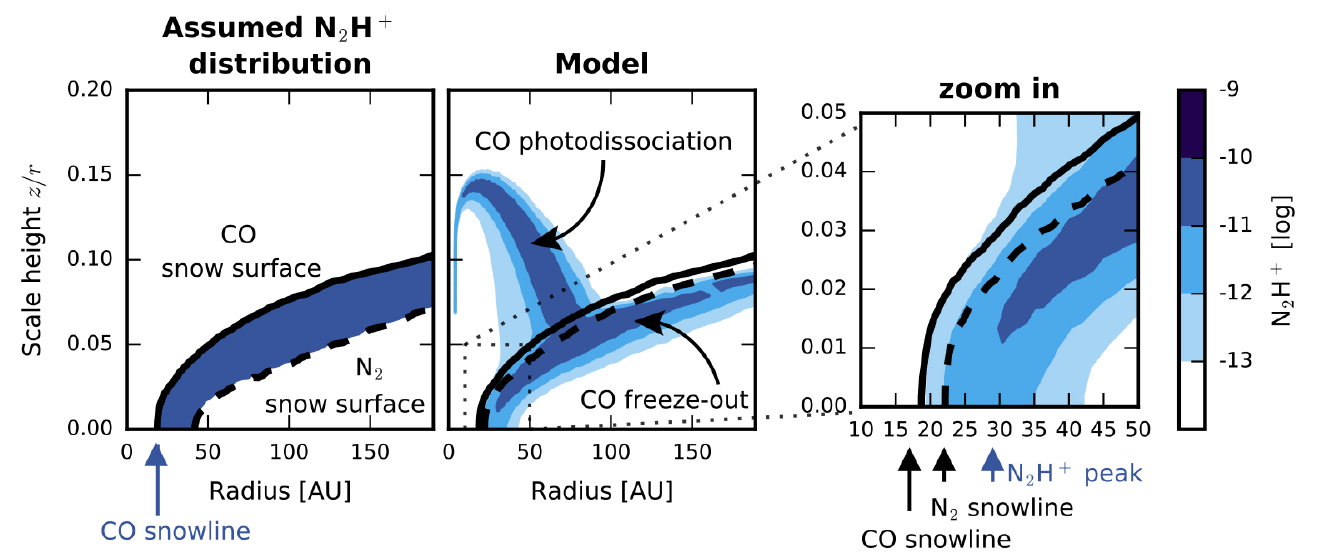} 
 \caption{Generally assumed N$_2$H$^+$ distribution in protoplanetary disks, i.e., between the CO and N$_2$ snow surfaces (\textit{left panel}) and the N$_2$H$^+$ distribution for TW~Hya as predicted by our simple chemical model (\textit{middle panel}). A zoom in on the region around the snowlines is shown in the \textit{right panel}. The solid line indicates the CO snow surface and the dashed line the N$_2$ snow surface. The crucial differences between the expectations and the model predictions are the location of N$_2$H$^+$ peak (at the CO snowline versus beyond the CO and N$_2$ snowlines, resp.), and the formation of N$_2$H$^+$ above the CO snow surface in the model.}
 \label{fig:COsnowline}
\end{center}
\end{figure}

An alternative approach is to use a molecule whose abundance is strongly affected by the freeze-out of CO. One such molecule is DCO$^+$, which forms via reaction of H$_2$D$^+$ and CO. Based on these chemical considerations, DCO$^+$ emission is expected to form a ring inside the CO snowline, where the CO abundance is low enough to enhance H$_2$D$^+$ formation, while there is still enough gaseous CO left to act as parent molecule for DCO$^+$. \cite[Mathews et al. (2013)]{Mathews2013} indeed observed ring-shaped DCO$^+$ emission toward HD 163296, but the outer radius was later shown not to correspond exactly to the CO snowline (\cite{Qi2015}); probably because DCO$^+$ can also form in warmer regions higher up in the disk from CH$_2$D$^+$ (\cite{Favre2015}). ALMA observations of six protoplanetary disks by \cite[Huang et al. (2017)]{Huang2017} corroborate that DCO$^+$ emission, while still tracing relatively cold gas, does not have a direct relation with the CO snowline in disks. 

Another molecule suggested to trace the CO snowline is N$_2$H$^+$. Formation of N$_2$H$^+$ occurs through proton transfer from H$_3^+$ to N$_2$, 
\begin{equation} \label{eq:N2H+formation}
\mathrm{N_2 + H_3^+} \rightarrow \mathrm{N_2H^+ + H_2},
\end{equation}
but is impeded when CO is present in the gas phase because CO competes with N$_2$ for reaction with H$_3^+$, 
\begin{equation} \label{eq:HCO+formation}
\mathrm{CO + H_3^+} \rightarrow \mathrm{HCO^+ + H_2}. 
\end{equation} 
In addition, reaction with gaseous CO is the dominant destruction route of N$_2$H$^+$: 
\begin{equation} \label{eq:N2H+destruction}
  \mathrm{N_2H^+ + CO} \rightarrow \mathrm{HCO^+ + N_2}.
\end{equation}
N$_2$H$^+$ is therefore expected to be abundant only beyond the CO snowline, where CO is depleted from the gas phase. A CO snowline location has indeed been derived from N$_2$H$^+$ emission for the disks around TW Hya and HD 163296 (\cite[Qi et al. 2013, 2015]{Qi2013}). 

However, a simple chemical model incorporating the main reactions for N$_2$H$^+$ (Eqs. \ref{eq:N2H+formation}--\ref{eq:N2H+destruction}) in addition to freeze out, thermal desorption and photodissociation of CO and N$_2$, combined with radiative transfer modeling, shows that the relationship between N$_2$H$^+$ and the CO snowline is more complicated (\cite{vantHoff2017}). First, the N$_2$H$^+$ abundance peaks at temperatures slightly below the CO freeze-out temperature, instead of directly at the CO snowline (see Fig.~\ref{fig:COsnowline}), as also found by \cite[Aikawa et al. (2015)]{Aikawa2015}. The snowline marks the radius where 50\% of the CO is present in the gas phase and 50\% is frozen out. Apparently, this reduction in gaseous CO is not yet enough to diminish the N$_2$H$^+$ destruction. Second, N$_2$H$^+$ can be formed higher up in the disk, above the layer where CO is frozen out, due to a small difference in the photodissociation rates for CO and N$_2$ (see Fig.~\ref{fig:COsnowline}). The slightly higher rate for CO creates another region where N$_2$ is still present in the gas phase, while CO is not. This ``surface'' layer of N$_2$H$^+$ can contribute significantly to the emission, shifting the N$_2$H$^+$ emission peak to larger radii, away from the CO snowline. N$_2$H$^+$ emission alone then merely provides an upper limit for the snowline location. 

\begin{figure}[t]
\begin{center}
 \includegraphics[width=0.4\linewidth]{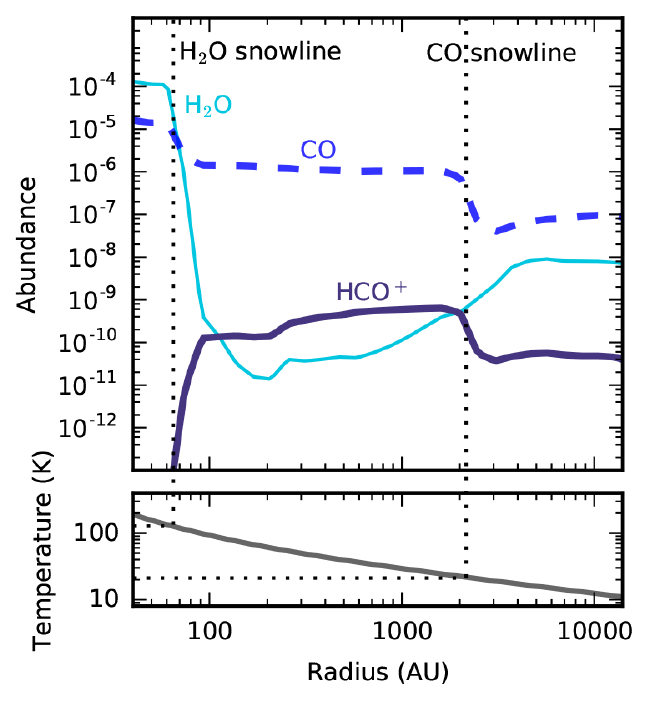} 
 \caption{Temperature profile for the NGC1333 IRAS2A envelope from \cite[Kristensen et al. (2012)]{Kristensen2012} (\textit{bottom panel}), and the corresponding H$_2$O (thin solid line), CO (dashed line) and HCO$^+$ (thick solid line) abundances predicted by the three-phase astrochemical model GRAINOBLE (\textit{top panel}). The vertical dotted lines mark the H$_2$O and CO snowlines around 100 K and 20 K, resp.}
 \label{fig:HCO+chemistry}
\end{center}
\end{figure}

Applying our modeling approach to the TW~Hya disk, using the physical model from \cite[Kama et al. (2016)]{Kama2016}, we derive a CO snowline location of $\sim$18~AU, instead of the $\sim$30~AU suggested by \cite[Qi et al. (2013)]{Qi2013}. The latter authors fitted a radial N$_2$H$^+$ column density profile to the emission with a steep rise at the CO snowline. Our outcome is consistent with the results from \cite[Schwarz et al. (2016)]{Schwarz2016} based on multiple $^{13}$CO and C$^{18}$O lines, and the $^{13}$C$^{18}$O analysis by \cite[Zhang et al. (2017)]{Zhang2017}. Deriving the CO snowline location from N$_2$H$^+$ emission is thus not as straightforward as was generally assumed, but, given a good physical model for the target disk, the location can be determined using simple chemistry in combination with radiative transfer modeling.




\section{The H$_2$O snowline}

The most important snowline is the water snowline, since the bulk of the oxygen budget and ice mass is in water ice. Unfortunately, this snowline is even more difficult to observe than that for CO. Because of the large binding energy of H$_2$O, water sublimates off the grains only at temperatures above $\sim$100~K. This means that the snowline is located a few AU from the star in protoplanetary disks, that is, $\sim$0.01$^{\prime\prime}$ in the nearest star-forming region Taurus. High angular resolution is thus required to observe it. Furthermore, except for the H$_2$O line at 183 GHz, the only thermal water lines observable from the ground are those from the less abundant isotopologue H$_2^{18}$O. As such, even ALMA will have great difficulty to detect the water snowline in protoplanetary disks. Cold water lines ($<$100~K) have been detected from space toward TW Hya with \textit{Herschel} (\cite{Hogerheijde2011}), but the \textit{Herschel} beam is too large to resolve the snowline. So far, only for the disk around V883 Ori a snowline location has been reported, but this was inferred from a steep drop in the dust optical depth (\cite{Cieza2016}). 

\begin{figure}[t]
\begin{center}
 \includegraphics[width=\linewidth]{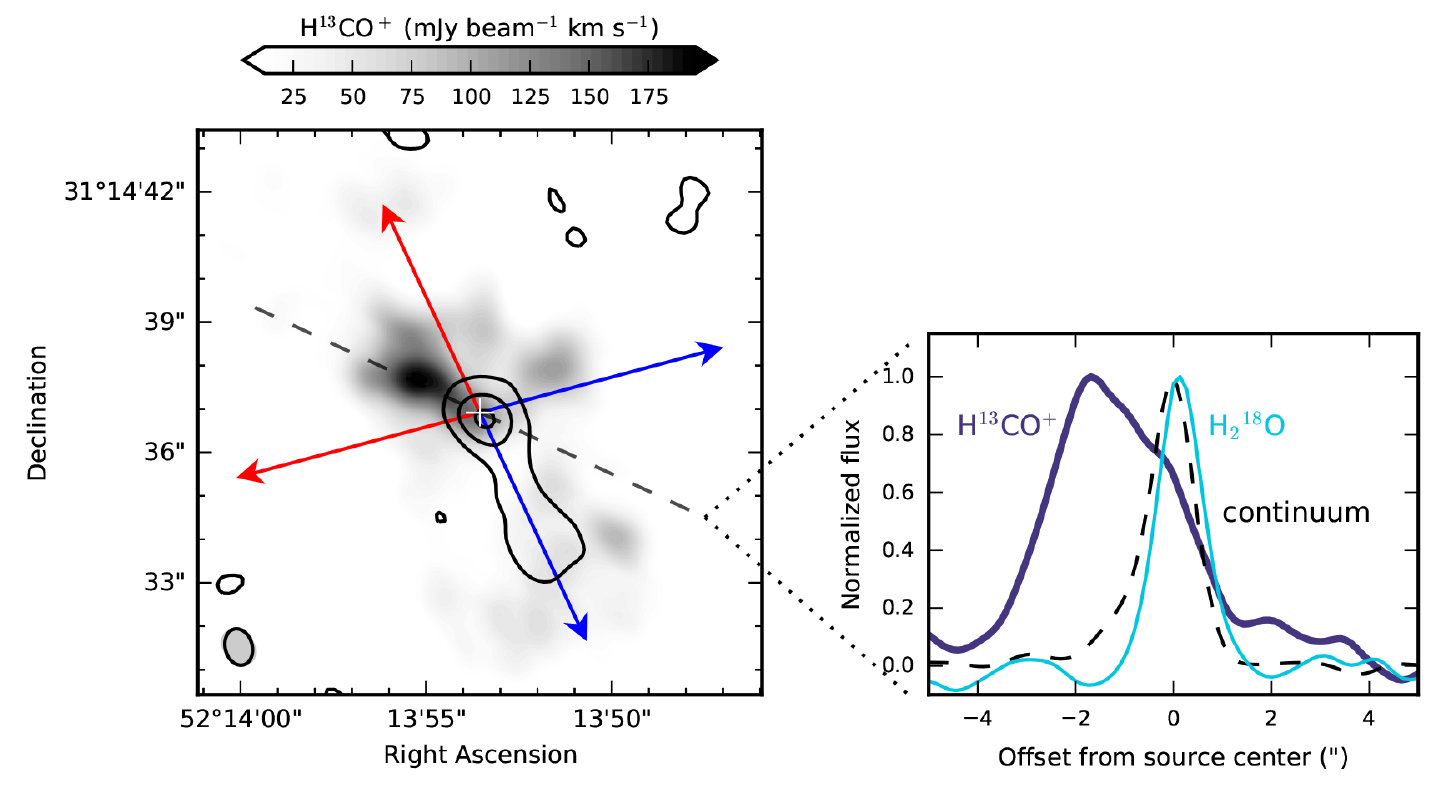} 
 \caption{Integrated continuum subtracted intensity map for the H$^{13}$CO$^+$ $J=3-2$ transition (grey scale) toward IRAS2A with the H$_2^{18}$O $3_{1,3}-2_{2,0}$ transition overlaid in contours (\textit{left panel}). The H$_2^{18}$O contours are shown at 8.2 (1$\sigma$) $\times$ [3, 15, 35] mJy~beam$^{-1}$~km~s$^{-1}$. The position of the continuum peak is marked by a white cross and the outflow axes by arrows. Both observations have a similar beam size (depicted in the lower left corner). The \textit{right panel} shows the integrated intensities for H$^{13}$CO$^+$ (thick solid line), H$_2^{18}$O (thin solid line) and the continuum (dashed line) along the dashed line, normalized to their maximum value.}
 \label{fig:H13CO+observations}
\end{center}
\end{figure}

The best way to locate the water snowline may therefore be by applying the same strategy as done for CO, that is, using a chemical tracer. The best candidate to trace the water snowline is HCO$^+$, because gaseous H$_2$O is its most abundant destroyer (\cite{Phillips1992}; \cite{Bergin1998}). A strong decline in HCO$^+$ abundance is thus expected when H$_2$O desorbs from the dust grains. This is corroborated by the results from chemical models. Figure~\ref{fig:HCO+chemistry} shows the outcome of the three-phase astrochemical code GRAINOBLE (\cite{Taquet2014}) for the 1D temperature and density structure of the envelope around NGC1333 IRAS2A (\cite{Kristensen2012}). Three-phase models consider reactions for gas-phase species, reactions for species on the ice surface, and reactions for bulk ice species. As expected, the HCO$^+$ and H$_2$O abundances show a strong anticorrelation. Hints of this anticorrelation are seen in observations towards the protostar IRAS 15398-3359, which show ring-shaped H$^{13}$CO$^+$ emission surrounding CH$_3$OH emission, another grain-surface molecule with a similar snowline location as water (\cite{Jorgensen2013}). The non-detection of the high excitation H$_2^{18}$O $4_{1,4}-3_{2,1}$ line prevented unambigious confirmation, but the detected HDO emission, although more complex, is consistent with the H$_2$O-HCO$^+$ anticorrelation scenario (\cite{Bjerkeli2016}). 

\begin{figure}[t]
\begin{center}
 \includegraphics[width=\linewidth]{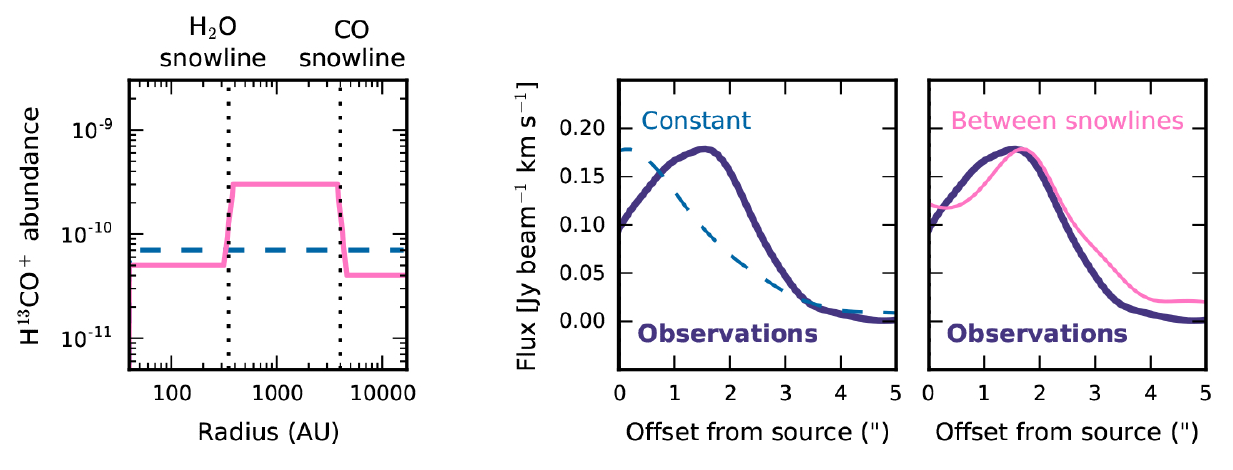} 
 \caption{Different H$^{13}$CO$^+$ abundance profiles (\textit{left panel}) and the resulting simulated integrated emission along the northeastern part of the radial cut shown in Fig.~\ref{fig:H13CO+observations} (\textit{right panels}). The vertical dotted lines in the \textit{left panel} mark the H$_2$O and CO snowlines around 100 K and 20 K, resp. The observed H$^{13}$CO$^+$ emission is shown with the thick solid lines in the \textit{right panels}. }
 \label{fig:H13CO+models}
\end{center}
\end{figure}

Protostellar envelopes, like IRAS 15398-3359, are the best sources to establish whether HCO$^+$ is a good snowline tracer: the water snowline is located further away from the star than in disks (10s--100s AU rather than a few AU; \cite{Harsono2015}; \cite{Cieza2016}), and compact warm water has already been observed toward four sources (\cite{Jorgensen2010}; \cite{Persson2012}, 2013; \cite{Taquet2013}). The only thing lacking are thus HCO$^+$ observations. We have therefore observed the optically thin isotopologue H$^{13}$CO$^+$ toward the Class 0 protostar NGC1333 IRAS2A using NOEMA (\cite{vantHoff2017b}). Comparison with the H$_2^{18}$O emission presented by \cite[Persson et al. (2012)]{Persson2014} shows that while H$_2^{18}$O peaks on source, H$^{13}$CO$^+$ has its emission peak $\sim$~2$^{\prime\prime}$ off source (see Fig.~\ref{fig:H13CO+observations}). As a first analysis, we performed a 1D radiative transfer calculation with Ratran (\cite{Hogerheijde2000}), using the 1D temperature and density structure derived by  \cite[Kristensen et al. (2012)]{Kristensen2012} from DUSTY modeling (\cite{Ivezic1997}) of the continuum emission. A simple parametrized abundance profile for H$^{13}$CO$^+$ with sharp decreases inside the H$_2$O snowline and outside the CO snowline, as predicted by a full chemical model, can reproduce the observed location of the emission peak (see Fig.~\ref{fig:H13CO+models}). The H$^{13}$CO$^+$ abundance drops outside the CO snowline because its parent molecule, CO, is frozen out. The temperature, however, has to be increased by a factor of $\sim$2 to reproduce the observed peak location with a drop in H$^{13}$CO$^+$ at 100~K. A constant H$^{13}$CO$^+$ abundance at all radii, on the other hand, produces emission peaking on source, unlike observed. These results suggest that water and HCO$^+$ are indeed anticorrelated, and provide a proof of concept that the optically thin isotopologue H$^{13}$CO$^+$ can be used to trace the water snowline.

\section{Summary and outlook}

Due to its close proximity to the central star and the fact that only the less abundant isotopologue H$_2^{18}$O can be observed from the ground, both high angular resolution and high sensitivity are required to observe the water snowline. Chemical imaging may therefore be the only way to locate this snowline in protoplanetary disks. This approach has proven useful for the CO snowline, although simple chemical considerations have to be taken into account when deriving the CO freeze-out radius from N$_2$H$^+$ emission. The best candidate to image the water snowline is HCO$^+$, because its main destructor is gaseous H$_2$O. The first observations of both H$^{13}$CO$^+$ and H$_2^{18}$O toward the protostellar envelope of NGC1333 IRAS2A show that these molecules are indeed spatially anti-correlated. This suggest that H$^{13}$CO$^+$ may be used as a tracer of the water snowline in protoplanetary disks.



\section*{Acknowledgments}

I would like to acknowledge and thank Catherine Walsh, Mihkel Kama, Stefano Facchini, Magnus Persson, Daniel Harsono, Vianney Taquet, Jes J{\o}rgensen, Ruud Visser, Edwin Bergin and Ewine van Dishoeck for their contributions to this work, and support from a Huygens fellowship from Leiden University.



\begin{thebibliography}{}

\bibitem[Aikawa et al. 2015]{Aikawa2015}
{Aikawa, Y., Furya, K., Nomura, H., \& Qi, C.} 2015,
\textit{ApJ}, 807, 120

\bibitem[Ansdell et al. 2016]{Ansdell2016}
{Ansdell, M., Williams, J.P., van der Marel, N.} 2016,
\textit{ApJ}, 828, 46

\bibitem[Bergin et al. 1998]{Bergin1998}
{Bergin, E.A., Melnick, G.J., \& Neufeld, D.A.} 1998, 
\textit{ApJ}, 499, 777

\bibitem[Bjerkeli et al. 2016]{Bjerkeli2016}
{Bjerkeli, P., J{\o}rgensen, J.K., Bergin, E.A., \etal} 2016,
\textit{A\&A}, 595, A39

\bibitem[Cieza et al. 2016]{Cieza2016}
{Cieza, L.A., Casassus, S., Tobin, J., \etal} 2016, 
\textit{Nature}, 535, 258

\bibitem[Eistrup et al. 2016]{Eistrup2016}
{Eistrup, C., Walsh, C., \& van Dishoeck, E.F.} 2016,
\textit{A\&A}, 595, A83

\bibitem[Favre et al. 2015]{Favre2015}
{Favre, C., Bergin, E.A., Cleeves, L.I., Hersant, F., Qi, C., \& Aikawa, Y.} 2015, 
\textit{ApJL}, 802, L23

\bibitem[Herbst \& van Dishoeck 2009]{Herbst2009}
{Herbst, E., \& van Dishoeck, E.F.} 2009,
\textit{ARA\&A}, 47, 427

\bibitem[Harsono et al. 2015]{Harsono2015}
{Harsono, D., Bruderer, S., \& van Dishoeck, E.F.} 2015,
\textit{A\&A}, 582, A41

\bibitem[Hogerheijde \& van der Tak 2000]{Hogerheijde2000}
{Hogerheijde, M.R., \& van der Tak, F.F.S.} 2000,
\textit{A\&A}, 362, 697

\bibitem[Hogerheijde et al. 2011]{Hogerheijde2011}
{Hogerheijde, M.R., Bergin, E.A., Brinch, E.A., \etal} 2011,
\textit{Science}, 334, 338

\bibitem[Huang et al. 2017]{Huang2017}
{Huang, J., {\"O}berg, K.I., Qi, C., \etal} 2017,
\textit{ApJ}, 835, 231

\bibitem[Ivezi{\'c} \& Elitzur 1997]{Ivezic1997}
{Ivezi{\'c}, Z., \& Elitzur, M.} 1997, 
\textit{MNRAS}, 287, 799

\bibitem[J{\o}rgensen \& van Dishoeck 2010]{Jorgensen2010}
{J{\o}rgensen, J.K., \& van Dishoeck, E.F.} 2010,
\textit{ApJL}, 710, L72

\bibitem[J{\o}rgensen et al. 2013]{Jorgensen2013}
{J{\o}rgensen, J.K., Visser, R., Sakai, N., \etal} 2013,
\textit{ApJL}, 779, L22

\bibitem[Kama et al. 2016]{Kama2016}
{Kama, M., Bruderer, S., van Dishoeck, E.F., \etal} 2016,
\textit{A\&A}, 592, A83

\bibitem[Kristensen et al. 2012]{Kristensen2012}
{Kristensen, L.E., van Dishoeck, E.F., Bergin, E.A., \etal} 2012, 
\textit{A\&A}, 542, A8

\bibitem[Madhusudhan et al. 2014]{Madhusudhan2014}
{Madhusudhan, N., Amin, M.A., \& Kennedy, G.M.} 2014,
\textit{ApJL}, 794, L12

\bibitem[Mathews et al. 2013]{Mathews2013}
{Mathews, G.S., Klaassen, P.D., Juh{\'a}sz, A., \etal} 2013,
\textit{A\&A}, 557, A132

\bibitem[Miotello et al. 2017]{Miotello2017}
{Miotello, A., van Dishoeck, E.F., Williams, J.P., \etal} 2017,
\textit{A\&A}, 599, A113

\bibitem[{\"O}berg et al. 2011]{Oberg2011}
{{\"O}berg, K.I., Murray-Clay, R., \& Bergin, E.A.} 2011, 
\textit{ApJL}, 743, L16

\bibitem[{\"O}berg \& Bergin 2016]{Oberg2016}
{{\"O}berg, K.I., \& Bergin, E.A.} 2016,
\textit{ApJL}, 831, L19

\bibitem[Persson et al. 2012]{Persson2012}
{Persson, M.V., J{\o}rgensen, J.K., \& van Dishoeck, E.F.} 2012, 
\textit{A\&A}, 541, A39

\bibitem[Persson et al. 2013]{Persson2013}
{Persson, M.V., J{\o}rgensen, J.K., \& van Dishoeck, E.F.} 2013,
\textit{A\&A}, 549, L3

\bibitem[Phillips et al. 1992]{Phillips1992}
{Phillips, T.G., van Dishoeck, E.F., \& Keene, J.} 1992,
\textit{ApJ}, 399, 533

\bibitem[Qi et al. 2013]{Qi2013}
{Qi, C., {\"O}berg, K.I., Wilner, D.J., \etal} 2013,
\textit{Science}, 341, 630

\bibitem[Qi et al. 2015]{Qi2015}
{Qi, C., {\"O}berg, K.I., Andrews, S.M., \etal} 2015,
\textit{ApJ}, 813, 128

\bibitem[Ros \& Johansen 2013]{Ros2013}
{Ros, K., \& Johansen, A.} 2013,
\textit{A\&A}, 552, A137

\bibitem[Schoonenberg \& Ormel 2017]{Schoonenberg2017}
{Schoonenberg, D., \& Ormel, C.W.} 2017,
\textit{A\&A}, 602, A21

\bibitem[Schwarz et al. 2016]{Schwarz2016}
{Schwarz, K., Bergin, E.A., Cleeves, L.I., \etal} 2016,
\textit{ApJ}, 823, 91

\bibitem[Stevenson \& Lunine 1988]{Stevenson1988} 
{Stevenson, D.J., \& Lunine, J.I.} 1988,
\textit{Icarus}, 75, 146

\bibitem[Taquet et al. 2013]{Taquet2013}
{Taquet, V., L{\'o}pez-Sepulcre, A., Ceccarelli, C., \etal} 2013,
\textit{ApJL}, 768, L29

\bibitem[Taquet et al. 2014]{Taquet2014}
{Taquet, V., Charnley, S.B., \& Sipil{\"a}, O.} 2014, 
\textit{ApJ}, 791, 1

\bibitem[van 't Hoff et al. 2017a]{vantHoff2017}
{van 't Hoff, M.L.R., Walsh, C., Kama, M., Facchini, S., \& van Dishoeck, E.F.} 2017a,
\textit{A\&A}, 599, A101 

\bibitem[van 't Hoff et al. 2017b]{vantHoff2017b}
{van 't Hoff, M.L.R., Persson, M.V., Harsono, D., \etal} 2017b, 
\textit{A\&A}, submitted

\bibitem[Walsh et al. 2015]{Walsh2015}
{Walsh, C., Nomura, H., \& van Dishoeck, E.F.} 2015,
\textit{A\&A}, 582, A88

\bibitem[Zhang et al. 2017]{Zhang2017}
{Zhang, K., Bergin, E.A., Blake, G.A., Cleeves, L.I., \& Schwarz K.R.} 2017
\textit{Nat. Astron.}, 1, 0130






\end{thebibliography}
\end{document}